\documentclass[twocolumn,showpacs,preprintnumbers,amsmath,amssymb]{revtex4}
\usepackage{graphicx}
\usepackage{amsmath}
\usepackage{amstext}
\usepackage{latexsym}
\usepackage{psfig}
\usepackage{amsfonts}
\usepackage{bm}
\usepackage{amssymb}
\usepackage{amscd}
\begin{document}
\title{Quantum Key Distribution with Blind Polarization Bases}
\author{Won-Ho Kye and Chil-Min Kim}
\affiliation{National Creative Research Initiative Center for Controlling Optical Chaos, Pai-Chai University, Daejeon 302-735, Korea}
\author{M. S. Kim}
\affiliation{School of Mathematics and Physics, Queen's University, Belfast BT7 1NN, United Kingdom}
\author{Young-Jai Park}
\affiliation{Department of Physics, Sogang University, Seoul 121-742, Korea}
\date{\today}
\begin{abstract}
We propose a new quantum key distribution scheme that uses the blind
polarization basis. In our scheme the sender and the receiver share key
information by exchanging qubits with arbitrary polarization
angles without basis reconciliation.  As only
random polarizations are transmitted, our protocol is secure even when
a key is embedded in a not-so-weak coherent-state pulse.  We show its
security against the photon number splitting attack and the
impersonation attack.

\end{abstract}
\pacs{03.67.-a,03.67.Dd,03.67.Hk }
\maketitle
Quantum key distribution (QKD)
\cite{BB84,Gisin-review,Luet,Ekert,Grangier,B92}, whose security is
guaranteed by the laws of physics, has attracted widespread
attention as it is ultimately secure and does not require
computational and mathematical complexity unlike its classical
counterparts \cite{RSA}.  Since the seminal work of a QKD protocol (BB84) by
Bennett and Brassard \cite{BB84}, there have been theoretical
proposals, experimental realizations \cite{Gisin-review} and their
security proofs \cite{security}. 
Recently, the security was proved based on entanglement
\cite{Luet} for both the single-photon QKD \cite{BB84} and the
entangled-state QKD \cite{Ekert}. The
continuous-variable QKD \cite{Grangier}  
has also been proved to be a promising protocol to send secret keys with
high transmission rate.      

The QKD is one of the most promising applications of quantum
information science and the gap between theory and practice has
become narrower.  In practical single-photon QKD,
the source sometimes 
inevitably produces more than one photon at a time.  In this case,
the QKD is vulnerable to the photon number splitting (PNS) attack
in which an eavesdropper (Eve) splits photons from
the many-photon field and keeps them. Eve measures her photons when the bases
are announced via the classical channel. The PNS attack is not always the
best strategy, for example,  as the photon cloning attack can
sometimes be more powerful.  The two attacks were  compared in
\cite{PNS}. 

Recently Bostr\"om and Felder \cite{ping} came up with a conceptually
new type of quantum secret coding,
which allows information to be transmitted in a deterministic secure
way, based on entanglement and two-way communication.  This direct
quantum coding protocol, 
sometimes called the {\it ping-pong} protocol, has been extended
to single-photon implementation \cite{ping-polar} and its security
was extensively studied \cite{ping-attack}.  

In this Letter, we propose a new QKD scheme in which the basis
reconciliation via a classical channel is not necessary as an
advantage.  In this QKD
scheme, Alice chooses a random value of angle
$\theta$ and prepares a 
photon state with the polarization of that angle. Bob also chooses
another random value of angle $\phi$ and further rotates
the polarization direction of the received photon state by $\phi$ and then
returns  to Alice. Alice encodes the 
message by rotating the polarization angle by $\pm \pi/4$ after
compensating the angle by $-\theta$. Bob reads the photon state by measuring the
polarization, after compensating the angle by
$-\phi$.  Alice and Bob shall choose 
random angles, $\theta$ and $\phi$, for each transmission of qubits. 
This will be continued until the desired number of bits are created.

The important advantages of our protocol are manifold: 1) The
reconciliation of the polarization basis is not necessary.  The strong
point of our protocol is that the selected polarization angles
$\theta$ and $\theta+\phi$ are not
necessary to discuss with each other as 
Bob and Alice do not need to know the other's polarization basis.
Moreover, this may significantly increase the bit creation rate
compared to the  
BB84 protocol and, in an ideal case, enables the direct quantum
coding \cite{Lo}. 
2) In most other two-way
QKD's, the receiver sends a random qubit and the key sender randomly
decides one of the two sets of unitary 
operations on it. The unitary operation then becomes the key in the
two-way QKD.  A problem with this scheme is that the key may be
disclosed to an eavesdropper who 
sends a spy photon along with the receiver's random photon traveling
to the sender\cite{Boil}.  However, in our protocol, all codings are random,
which makes it robust against such the attack. 3) Our coding may be
implemented by laser pulses (the security is guaranteed even for
relatively high-intensity laser pulses).  This advantage is due to the
fact that the polarizations
are completely arbitrary which makes our protocol resistant to both
the PNS attack and the attack based on two photon interference
\cite{Hong}.  Meanwhile, one of the practical difficulties of the
implementation of 
polarization QKD is the random fluctuation of the polarization due to
birefringence in the fiber.  However, as
Muller {\em  et al.} \cite{Gisin} pointed out, the time scale of the
random fluctuations in fibers is tens of minutes which is long enough
to enable polarization tracking to compensate them.  The rate of errors caused
by technical imperfections using today's technology is of the order of
a few percent under which our protocol is considered to be secure
\cite{Gisin-review} .   

{\it Protocol.-} The procedure for the proposed QKD is as follows:
\begin{itemize}
\item [(a.1)] \label{a.1}
Alice prepares a linearly polarized qubit in its initial state $|\psi_0
\rangle=|0 \rangle$, where $|0\rangle$ and $|1\rangle$ represent two
orthogonal polarizations of the qubit, and chooses a
random angle $\theta$.

\item [(a.2)] \label{a.2}
Alice rotates the polarization of the qubit by $\theta$
to bring the state of the qubit to  
$|\psi_1\rangle =\hat{U}_y(\theta)|\psi_0\rangle=\cos\theta|0\rangle-
\sin\theta|1\rangle$, where $\hat{U}_y(\theta)=\cos\theta\openone-
i\sin\theta\hat{\sigma}_y$ is the unitary operator which rotates the
polarization angle along the $y$ axis and $\hat{\sigma}_y$ is the
Pauli-$y$ operator. Alice sends the qubit to Bob.

\item [(a.3)] \label{a.3}
Bob chooses another random angle $\phi$ and rotates the
polarization of the received qubit by $\phi$; $|\psi_2\rangle
=\hat{U}_y(\phi)|\psi_1\rangle=\cos(\theta+\phi)|0\rangle-
\sin(\theta+\phi)|1\rangle$.  Bob sends the qubit back to Alice.

\item[(a.4)] \label{a.4} 
Alice rotates the polarization angle of the qubit by $-\theta$
and then encodes the message by further rotating the polarization angle
of $\pm\pi/4$; $|\psi_3\rangle=\hat{U}_y(\pm\pi/4)\hat{U}_y(-\theta)|\psi_2
\rangle$. Alice sends the qubit to Bob.  (Alice and Bob
have predetermined that $\pi/4$ is, say, ``0'' and $-\pi/4$ is ``1''.)

\item[(a.5)] \label{a.5} 
Bob measures the polarization after
rotating the polarization by $-\phi$; $|\psi_4\rangle =
\hat{U}_y(-\phi)|\psi_3\rangle=\hat{U}_y(\pm\pi/4)|\psi_0\rangle$.  
$\hat{U}_y(+\pi/4)|\psi_0\rangle$ and
$\hat{U}_y(-\pi/4)|\psi_0\rangle$ are orthogonal to each other, which
enables Bob to read the keys precisely.
\end{itemize}

By repeating the above protocol $k$ times with different
random phases, Alice and Bob share $k$ bits of information. In order
to verify the  integrity of the shared key, the convention is to use a
public channel to 
reveal some part of key bits \cite{BB84, ping-polar}. That kind of
verification method has two weak points. 1) It usually degrades the
efficiency of the key distribution. 2) It does not guarantee the
integrity of the remaining key bits.  In order to overcome those
problems, we shall use the one-way hash function \cite{Hash} for
checking the integrity of the shared key bits. 
\begin{itemize}
\item[(a.6)] \label{a.6}
Alice announces the
one-way hash function $H$ via a classical channel.  Alice and Bob
evaluate the hash values, $h_a=H(k_a)$ and $h_b=H(k_b)$ respectively,
where $k_a$ and $k_b$ are shared keys in Alice and Bob.  If $h_a=h_b$,
they keep the shared keys, otherwise, they abolish the keys and start
the process again from (a.1).
\end{itemize}
In (a.6) the difference between $h_a$ and $h_b$ implies that Alice and
Bob do not share the exactly same keys.  This is due to imperfection in
the transmission or to Eve who intervened between Alice and Bob.  
Figure 1 shows the sketch of the experimental setup.  

\begin{figure}
\rotatebox[origin=c]{0}{\includegraphics[width=8.5cm]{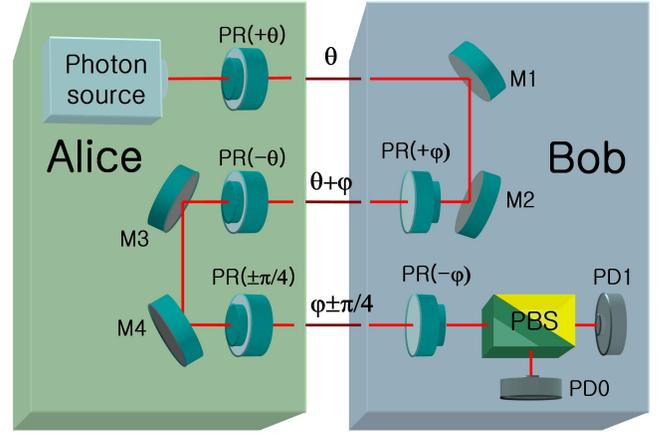}}
\caption{Schematic diagram for the experimental setup. PR:
  Polarization Rotator; M1, M2, M3, and M4: Mirrors; PD0 and PD1: 
        Photodetectors; PBS: Polarization Beam Splitter. 
}
\label{figure2}
\end{figure}

{\it Security.-} 
For a perfect channel with single-photon keys, it is obvious that
the eavesdropper cannot obtain much information by the
intercept-and-resend attack as all three signal transmissions are of random
polarizations.  In our protocol, the random polarizations lie on the
equator of the Poincar\'e sphere.  The optimum estimation of
a single-photon qubit in this case gives the fidelity 3/4 \cite{Buzek} where
the fidelity is one when the estimation is perfect or zero when the
original state is orthogonal to the estimation. (In this Letter, we
will consider the fidelity as the amount of information as in
\cite{Buzek,Popescu}.)  The interference can be easily noticed during
the protocol (a.6).  Another
possible attack is that Eve sends a spy pulse to Bob to find the
random value $\phi$ of Bob's operation.  In order to do so, Eve
should perform quantum tomography which requires an intense spy pulse.
This should be able to be noticed easily.  
 
A single-photon QKD is not very economical not only because
it is difficult to have a reliable single-photon source but also
because photons can easily be lost due to imperfect channel
efficiency.  In particular, our protocol has to travel three times
between Alice and Bob so the loss may not be negligible.  If a key is
encoded on a coherent-state pulse, the protocol can be vulnerable
against the PNS attack.  We thus examine the security of our protocol
against the PNS attack.  A coherent-state pulse of its amplitude
$\alpha$ ($\alpha\in\mathrm{I\!R}$) is $
|\alpha\rangle=\exp(-\alpha^2/2)\sum_{n=0}^{\infty}
\alpha^n/\sqrt{n!} |n\rangle
$
where $|n\rangle$ is the photon-number eigenstate.  The mean photon
number of the pulse is $\alpha^2$.  Let us assume that the channel
efficiency is $\eta$ for one trip either from Alice to Bob
or vice versa.  The amplitude then reduces to $\eta\alpha$ from $\alpha$.
For convenience, $\eta$ is taken to be real throughout the Letter.

(Attack 1) As usual, we assume that Eve is so superior that
her action is limited only by the laws of physics.  She replaces
the lossy channel by a perfect one and puts a beam splitter of the amplitude
transmittivity $\eta$ in the middle.  The reflected field, which is a
coherent state with its amplitude $\sqrt{1-\eta^2}\alpha$, will be the
source of information to Eve.  In the protocol (a.1)-(a.5),
the information transmitted between Alice and Bob is of random
polarization.  From earlier works \cite{Buzek,Popescu}, we know that
the maximum information one can obtain from a set of identically
prepared qubits whose polarization is completely unknown, depends on
the number of qubits.  Massar and Popescu \cite{Popescu} found that
the maximum amount of information which can be extracted from $n$ identical
spin-1/2 particles is $I(n) =(n+1)/(n+2)$ when the particle lies
at any point on the Poincar\'e sphere. However, in our protocol, the photon
polarizations lie on the equator of the Poincar\'e sphere. In this case,
Derka {\it et al}. \cite{Buzek} found that the optimum 
state estimation from $n$ qubits gives the maximal
mean fidelity
\begin{equation}
I(n)={1 \over 2}+\frac{1}{2^{n+1}}\sum_{\ell=0}^{n-1}\sqrt{
  \begin{pmatrix} n \\ \ell \end{pmatrix} \begin{pmatrix} n \\ \ell+1
  \end{pmatrix}}.
\label{info-2}
\end{equation}

Let us first consider the maximum information Eve can get from the
Alice$\rightarrow$Bob channel in (a.2).  The probability $P(n)$ of
there being $n$ photons in the coherent state
$|\sqrt{1-\eta^2}\alpha\rangle$ is 
\begin{equation}
P_{a.2}(n)=\exp[-(1-\eta^2)\alpha^2]\frac{[(1-\eta^2)\alpha^2]^n}{n!}.
\label{photon-number}
\end{equation}
Then the maximum amount of information Eve can
get from the channel in (a.2) is $I_{a.2}= \sum_{n=0}^\infty
P_{a.2}(n)I(n)$.  Eve has to take the PNS attack on the
Bob$\rightarrow$Alice channel in (a.3).  As Bob has received the
attenuated coherent state $|\eta\alpha\rangle$, the amplitude of Eve's
state is $\eta(1-\eta^2)^{1/2}\alpha$.  Similarly, the amplitude of
Eve's state from tapping the Alice$\rightarrow$Bob channel in
(a.4) is $\eta^2(1-\eta^2)^{1/2}\alpha$.  We calculate
the maximum information $I_{a.3}$ and $I_{a.4}$.

It is obvious \cite{ping-polar} that the overall maximum information
Eve can obtain is bounded by 
$I_E=\min(I_{a.2}, I_{a.3}, I_{a.4})$, which is plotted in Figs.~2
(solid lines) for various cases.  For the realistic channel
efficiency $\eta^2=0.5$, the pulse of its amplitude $\alpha=2.83$
gives the average number of photons delivered to Bob about 1 after the
process (a.4).  Figure 2 (a) shows that Eve's information in this case is
bounded by about 0.7,  while Alice and Bob always share
the perfect information, {\it i.e.} the amount of information $I_{AB}$
between Alice and Bob is unity.  Even though the probability of not
having a photon is about 36.8\% for a coherent-state
pulse of its amplitude 1, as Alice and Bob discard the empty qubits,
this should not lower the shared information.
      
As $\alpha$ gets larger, we see that
Eve's information becomes unity in Fig.~2 (a).  It is known that the rate
of the secure key depends on the difference between $I_{AB}$ and $I_E$ 
\cite{mutual}. As $\alpha$ gets larger the rate will decrease. Another
interesting result seen in Fig.~2 (b) is that the maximum bound for
Eve's information does not necessarily grow as the channel becomes
less efficient.  Eve's information is 
bounded by the minimum of $I_{a.2}$, $I_{a.3}$ and $I_{a.4}$ which
depend on the intensities of the qubit pulses during (a.2), (a.3) and
(a.4), respectively.  The intensity of the qubit pulse decreases as
the number of its laps between Alice and Bob 
increases.  $I_E$ is thus determined by $I_{a.4}$, 
proportional to the intensity of the qubit pulse,
$(1-\eta^2)\eta^4\alpha^2$ which maximizes at $\eta^2=2/3$.  When
$\eta$ is small, as many 
photons are taken out during the initial
transmission (a.2), there remain too few photons to give lesser
information in the later stage (a.4). The
information bound grows as $\eta$ grows but it
eventually converges to $I_E= 0.5$ as $\eta^2 
\rightarrow 1$, shown in Fig.~2 (b).  

(Attack 2) If Alice and Bob do not randomly check the intensities at
(a.2) and (a.3), Eve only needs to make sure that the final amplitude,
which Bob measures at (a.4), is $\eta^3\alpha$.  Then $I_E$
is optimized when Eve extracts the 
same amount of information at each step of (a.2), (a.3) and (a.4)
keeping the final amplitude which is measured by Bob.  In this case,
$I_E=I_{a.2}=I_{a.3}=I_{a.4}$ and the amplitude of Eve's field is always
$\sqrt{(1-\eta^6)/3}~\alpha$.  The amount of information is
plotted in dotted lines (Fig.~2), where $I_E$ monotonously
grows as the channel becomes less efficient.  For $\eta^2=0.5$ and
$\alpha=2.83$, $I_E\approx0.83$.

\begin{figure}
\rotatebox[origin=c]{0}{\includegraphics[width=8.5cm]{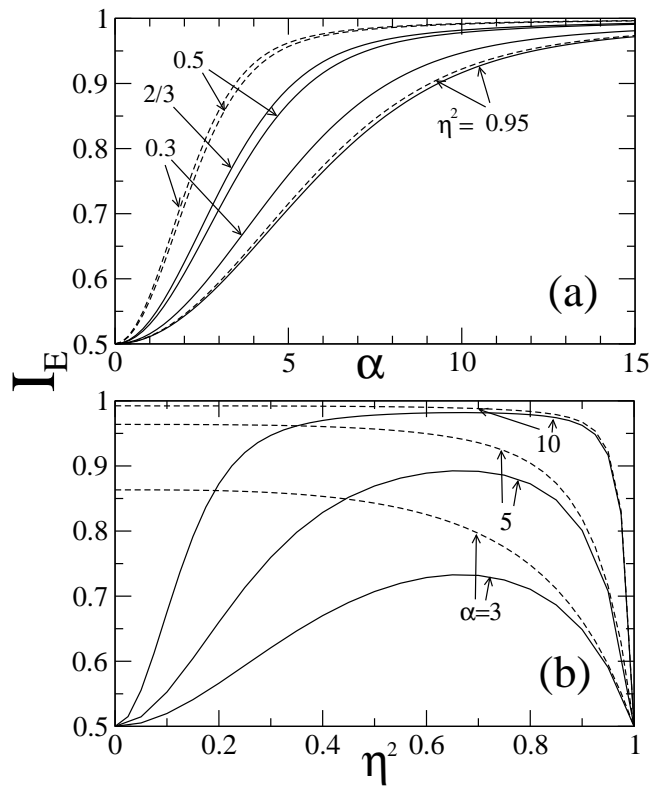}}
\caption{Maximum bound for Eve's information $I_E$ is plotted against
  the initial amplitude $\alpha$ of the coherent-state pulse
  (a). Various curves are
  according to various channel efficiencies 
  $\eta^2$ as shown along the curves.  In (b), $I_E$ is plotted against the
  channel efficiency for three values of the initial amplitude
  $\alpha$ of the pulse. Solid curves are for Attack 1 and dotted
  curves for Attack 2.    
}
\label{figure3}
\end{figure}
(Impersonation Attack) 
Our protocol is vulnerable to an impersonation attack \cite{im}
where Eve1 impersonates Bob to Alice reading the key then sends it to
Eve2.  Receiving the key, Eve2, who impersonates Alice to Bob, relays it
to Bob without being noticed.  We suggest to
slightly modify the protocol against this impersonation attack on the
quantum channel, leaving basic philosophy of the protocol the same.  Let us
consider that Alice, instead of one pulse, sends two coherent-state
pulses of the polarization
angles $\theta_1$ and $\theta_2$. Bob rotates the polarization
angles of the pulses by $\phi+(-1)^s\pi/4$ and $\phi+(-1)^{s\oplus
  1}\pi/4$, where the shuffling parameter $s\in\{0,1\}$ is randomly
chosen by Bob and $\oplus$ denotes addition modulo 2.  Receiving the two pulses of their polarization angles
$\theta_1+\phi+(-1)^s\pi/4$ 
and $\theta_2+\phi+(-1)^{s\oplus 1}\pi/4$,  Alice rotates the
polarization angles of the pulses by 
$-\theta_1+(-1)^k\pi/4$ and $-\theta_2+(-1)^k\pi/4$ respectively, where
$k\in\{0,1\}$ is the key value.
She blocks one of the qubits and sends the other to Bob. It is
important for Alice to delay the first pulse if it is let go, so that
impersonating Eve1 does not recognize which pulse was blocked.  Here, we
introduce the blocking factor $b$ to denote the case to let the
$b^{th}$ qubit go.  The qubit in state $|\psi^{sk}(b)\rangle$
will travel to Bob, where $|\psi^{00}(1)\rangle= |\psi^{10}(2)\rangle =
|\phi +\pi/2\rangle$ and $|\psi^{00}(2)\rangle
=|\psi^{10}(1)\rangle=|\phi\rangle$ for the key value $k=0$, and
$|\psi^{01}(1)\rangle =|\psi^{11}(2)\rangle=|\phi\rangle$ and
$|\psi^{01}(2)\rangle= |\psi^{11}(1)\rangle=|\phi-\pi/2\rangle$
for $k=1$. 
Upon receiving the qubit, Bob applies $\hat{U}_y(-\phi)$ on it and
measures the polarization. Depending on blocking as well as shuffling, he
obtains the measurement outcome $l^{sk}(b)=s\oplus k \oplus b$ as the
pre-key bit value.
Alice publicly announces her blocking factor $b$ for Bob to be able to
decode the original key bit $k$ by 
$k=s\oplus b\oplus l.$
Alice and Bob verify the shared key by exchanging the hash value of the key.
The shuffling parameter $s$ is Bob's private information, which
introduces additional random flipping of the key.  Eve's impersonation
without knowing $s$ value should induce errors in shared key with the
probability of 0.5 and be noticed during the comparison of the hash value. 
In implementation, two pulses of a key could be comfortably
manipulated for their separation of about 100nsec while the
distance between two keys may be in the order of 100$\mu$ sec. 
We have restricted our discussion only to the impersonation
attack on the quantum channel.  If such an attack is considered
for the public channel as well, an
authentication procedure has to be introduced.

{\it Remarks.}- 
We have considered a new QKD protocol which does not require
reconciliation of polarization bases.  All the operations are random
and independent, which makes the protocol robust against eavesdropping
attacks.  The protocol is secure even for a not-so-weak coherent-state
pulse, which may overlook a problem, a key has to travel
three times between Alice and Bob.
We have assumed for Alice and Bob to abolish the keys if hash
comparison was negative and also ignored possible noise.  The error rate
due to noise is currently about a few percentage, under which Eve's
information is not more than the information shared by Alice and Bob.
It should be possible to estimate a
security threshold and proceed with standard classical protocols to
distill a shorter secret key via privacy amplification.

{\it Acknowledgments}- We thank Professors H.-K. Lo, G. M. Palma and
W.Y. Hwang for comments.


\begin{thebibliography}{150}
\bibitem{Hash} A one-way hash function, $H(m)$, operates on an 
  arbitrary-length pre-image message $m$. It returns a fixed-length
  hash value (or message digest), $h$. The one-way hash function
  enables one to verify the 
  integrity of the shared message by exchanging the hash values
  through the public channel \cite{RSA}.
\end{thebibliography}
\end{document}